\documentstyle[12pt]{article}

\begin{document}

\begin{titlepage}
\null\vspace{-62pt}

\pagestyle{empty}
\begin{center}

\vspace{0.6truein} {\Large\bf   The quantum Yang-Mills theory}

\vspace{0.3in}
{\large Dimitrios Metaxas} \\
\vskip .4in
{\it Department of Physics,\\
National Technical University of Athens,\\
Zografou Campus, 15780 Athens, Greece\\
metaxas@mail.ntua.gr}\\

\vspace{.2in}
\centerline{\bf Abstract}

\baselineskip 18pt
\end{center}

In axiomatic quantum field theory, the postulate of the uniqueness of the vacuum (a pure vacuum state) is independent of the other axioms and equivalent to the cluster decomposition property.
The latter, however, implies a Coulomb or Yukawa attenuation of the interactions at growing distance, hence cannot accomodate 
the confining properties of the strong interaction.

The solution of the Yang-Mills quantum theory given previously, uses an auxiliary field to incorporate Gauss's law, and demonstrates the existence of two separate vacua, the perturbative and the confining vacuum, therefore a mixed vacuum state, deriving confinement, as well as the related, expected properties of the strong interaction.

The existence of multiple vacua is, in fact, expected by the axiomatic, algebraic quantum field theory, via the decomposition of the vacuum state to eigenspaces of the auxiliary field.
The general vacuum state is a mixed quantum state and the cluster decomposition property does not hold.
 Because of the energy density difference between the two vacua, the physics of the strong interactions does not admit a Lagrangian description.

I clarify the above remarks related to the previous solution of the Yang-Mills interaction, and conclude with some discussion, a criticism of a related mathematical problem, and some tentative comments regarding the spin-2 case.

\end{titlepage}
\newpage
\pagestyle{plain}
\setcounter{page}{1}
\newpage

\section{Introduction}

In \cite{dkm1}, I derived and applied some new Feynman diagram combinatorics, in order to incorporate the Gauss's law constraint in the perturbation expansion of the Yang-Mills theory, using an auxiliary, Lagrange multiplier field, $\lambda$.
This led to the derivation of a Coleman-Weinberg effective potential for $\lambda$, that appeared inverted in the resulting effective action, and admitted soliton solutions. However, the vacuum structure was not correctly identified in \cite{dkm1}, and the relation to actual properties of the Yang-Mills theory was not clear.

Eventually, in \cite{dkm2},  the vacuum  structure and stability were demonstrated, two distinct vacua emerged, which led to the proof of confinement, as well as various other expected properties of the strong interaction, and to a further confirmation of the often stressed importance of the vacuum state in deciphering a physical theory.

Here, I will use some well-known results of axiomatic, algebraic quantum field theory (AQFT) \cite{qft1, qft2, qft3}, in order to further justify the previous work. The discussion around Theorem 4.6 of \cite{qft3}, which I will mostly follow, is especially useful.
The relation of the uniqueness of the vacuum state to the clustering decomposition property (which was observed as early as in \cite{old}) will be described, and it will be shown that the presence of two separate vacua is related to the central decomposition of the GNS (Gelfand, Naimark, Segal) construction for the vacuum state, to eigenstates of the auxiliary field. 
The general vacuum state, then, is a mixed quantum state, described by a density matrix, and not a pure, unique vector, vacuum state, as it is often assumed in quantum field theory.

The formalism of AQFT that will be reviewed and applied here is especially constructed to ensure relativistic invariance of a quantum theory.
However, since the two vacua that emerge have different energy densities, there is no unique, Lorentz-invariant Lagrangian that can fully capture the physics of the strong interactions (if there was, one would be able to locally perturb one vacuum and lower the energy). Transitions between the two vacua happen only in a background of finite temperature or particle density, and the effective action derived in \cite{dkm1} can help elucidate many physical properties of the strong interactions, as demonstrated in \cite{dkm2}.

In Sec.~2, I start with the formalism of AQFT and in Sec.~3, I review some basic results of the solution of Yang-Mills theory, and apply the previous formalism. In Sec.~4,  I continue the description of the vacuum as a mixed state, and show the relation with the cluster decomposition.
In Sec.~5, I discuss possible relations with other works related to confinement, and in Sec.~6,
I conclude with some more comments, a criticism of the statement of a related mathematical problem, and some tentative remarks regarding the spin-2 interaction.

\section{Axiomatic, algebraic quantum field theory}

The formalism starts with an abstract algebra (that includes the physical observables), a physical state (a functional that also assigns expectation values to observables) and, via the GNS construction, gives them a representation on a concrete algebra of operators on a Hilbert space.

Here, by an abstract algebra, ${\cal A}$, I will denote a normed vector space over the complex numbers, ${\bf C}$, 
that is also an algebra (admits multiplication of its elements), 
admits conjugation (denoted here by $*$), and has a unity element with respect to multiplication. The norm and the conjugation operation have expected and compatible properties, for example, $(Q^*)^*=Q$, $ (Q_1 Q_2)^*=Q_2^* Q_1^*$, $||Q^* Q||=||Q||^2$, for any $Q$, $Q_1$, $Q_2$ in ${\cal A}$, and $||{\rm \bf I}||=1$.

For the purposes of this work, usually I will ignore issues of convergence and completion, and denote most units (algebraic, group theoretic, etc.) by ${\rm \bf I}$, unless these issues need specific clarification. Technically, ${\cal A}$ is a $C^*$-algebra with unity.

A state, ${\varphi}$, over ${\cal A}$, is a linear functional that assigns a complex number to each element, $Q$, of ${\cal A}$, and is positive ($\varphi(Q^* Q) \geq 0$) and normalised:
\begin{equation}
||\varphi|| \equiv \sup \{|\varphi(Q)|, Q \in {\cal A}, ||Q||\leq1 \}=1.
\end{equation}

Realistic (quantum, probabilistic) physical theories have physical observables, that are self-conjugate elements of the abstract algebra, ${\cal A}$, and the physical system under consideration is described by a state, $\varphi$, that bestowes the various expectation values.

The transition to the usual formalism of quantum mechanics is enabled by the GNS construction, which gives a representation of the aforementioned abstract algebra to a concrete algebra, that is,
 a subalgebra of ${\cal B}(H)$, the set of bounded linear operators on a Hilbert space, $H$, with the operation of the adjoint as conjugation (also denoted by $*$) and obviously also including the unity operator (technically, a von Neumann algebra).

Namely, for any state, $\varphi$, over an abstract algebra, ${\cal A}$, there exist a Hilbert space $H_{\varphi}$, 
a unit vector $\Omega_\varphi$ in $H_{\varphi}$, and a representation $\pi_\varphi$ of ${\cal A}$ 
on $H_{\varphi}$, that is,
$\pi_\varphi(Q) \in {\cal B}(H_{\varphi})$,
such that
\begin{equation}
\varphi(Q)= <\Omega_\varphi, \pi_\varphi(Q) \,\Omega_\varphi >,
\label{gns}
\end{equation}
where $<  \,,>$ denotes the inner product of the GNS Hilbert space, and
\begin{equation}
\{\pi_\varphi(Q) \,\Omega_\varphi , Q\in {\cal A}\}
\end{equation}
is dense in $H_{\varphi}$, that is, $\Omega_\varphi$ is a cyclic vector (any vector of  $H_{\varphi}$ can be 
approximated by $\pi_\varphi(Q) \,\Omega_\varphi $).

Note that the construction is specific to the state, $\varphi$, and applies to any state and physical theory.
Most crucially, however, note that the GNS representation is not necessarily irreducible:

For any subset, $S$, of ${\cal B}(H)$, denote by $S'$ the commutant of $S$, that is, the set of all elements of 
${\cal B}(H)$ that commute with every element of $S$. The center of $S$ is $S \cap S'$,  the set of all elements of $S$ that commute with every element of $S$.

Then, the GNS representation is irreducible if the center of $\pi_\varphi({\cal A})$ is trivial, consisting of  multiples of the identity operator.
Otherwise it is decomposed 
as a direct sum of eigenspaces of the operators of the center, and the state, $\varphi$, is mixed (described by a density matrix).
So, irreducibility or reducibility of the GNS construction is equivalent to whether the original state is pure or mixed.

The notation of (\ref{gns}) may lead to the confusion that this is a pure, vector state, but this is not the case, since one, obviously, could have started with a mixed state in the first place.

In the case of a non-trivial center, the resulting Hilbert space is decomposed as a direct sum $\oplus H_i$, the representation as 
$\oplus \pi_i$, the unit, cyclic vector as $\oplus c_i \Omega_i$, with complex coefficients $c_i,\, \sum_i |c_i|^2=1$,
and the original state is a mixed state, given by a density matrix
$\varphi = \sum _i \,\, |c_i|^2 |\Omega_i><\Omega_i|$.

This will be used and explained in more detail in later Sections, where it will be shown that the Yang-Mills vacuum state is a mixed state.

Like most fundamental results, the proof of the GNS construction is rather simple: the idea is to define the set
\begin{equation}
{\rm ker}\,\varphi =\{A\in {\cal A}: \varphi(A^* A) =0\},
\label{ker}
\end{equation}
use the Cauchy-Schwarz inequality (which holds generally)
\begin{equation}
|\varphi(A^* B)|^2 \leq \varphi (A^* A) \varphi(B^* B), \,\,\, \forall A, B \in {\cal A},
\end{equation}
to show that it is a left ideal ($\forall C \in {\cal A}, A \in {\rm ker} \,\varphi, C A \in {\rm ker}\,\varphi$),
and eventually, with the necessary completion procedures, construct the Hilbert space as a quotient space, ${\cal A}/{\rm ker}\,\varphi$,
with inner product of the equivalence classes of $A$ and $B$ given by $\varphi(A^* B)$.
 
Symmetries can also be naturally implemented in the algebraic formalism: to each element, $g$, of a symmetry group $G$, there corresponds an automorphism, $\alpha_g$, of ${\cal A}$, and if the state, $\varphi$, is invariant under $g$, that is, if
$\varphi(\alpha_g Q) = \varphi(Q)$ for all $Q \in {\cal A}$, then there exists a unitary operator, $U_\varphi (g)$ on the cyclic representation space $H_\varphi$ associated with $\varphi$, such that
$U_\varphi (g) \Omega_\varphi = \Omega_\varphi$ and 
$\pi_\varphi(\alpha_g (Q))=U_\varphi(g) \pi_\varphi (Q) U^*_\varphi (g)$.

In particular, we will consider the relativistic symmetry of the four-dimensional Minkowski spacetime, $M={\rm \bf R}^4$,
with metric $\eta^{\mu\nu}=(+---)$,  4-vectors $x=x^\mu =(x^0, \vec{x})$, $x\cdot y = x^0 y^0 - \vec{x} \vec{y}$.
As usual, $x$ will be called timelike, lightlike, and spacelike, depending on whether $x^2 =x \cdot x$ is positive, zero, or negative, respectively.

The causal complement, $D'$, of a subset $D$ of $M$, is the set of all points that have no causal relation with points of $D$,
$S'=\{x\in M :x-y \,\, {\rm spacelike}\,\, \forall y\in S\}$

The open future cone, $V_+=\{ x: x^2>0, x^0 >0\}$ and the closed future cone, 
$\bar{V}_+=\{ x: x^2 \ge 0, x^0  \ge 0\}$ and the similarly defined open past and closed past cone, are also important in describing the causal structure of the theory. The latter is preserved by the group of Poincare transformations, with elements 
$(\Lambda, a)$, where $\Lambda$ denotes a Lorentz transformation, and $a=a^\mu$ a spacetime translation.
Usually, one considers the group ${\cal P}_+^\uparrow$, where $\Lambda$ is in the connected component of the identity of the Lorentz group (proper, orthochronous Lorentz transformations).

A local, relativistic, quantum field theory, involves the abstract algebra of operators, ${\cal A}$, that is the union
of the abstract algebras ${\cal A}(D)$ for every bounded spacetime domain $D\subset M$. The observables of ${\cal A}(D)$
are the physical quantities measured by an observer within the space and time limits of $D$.
They obey the axioms:

(i) If $D_1 \subset D_2$ then ${\cal A}(D_1) \subset {\cal A}(D_2)$ (monotonicity: if an observable is measured in $D_1$ it can be measured in $D_2$).

(ii) If $D_1 \subset D'_2$ then ${\cal A}(D_1) \subset {\cal A}(D_2)'$ (causality: the  observables of two regions that are spacelike separated commute).

(iii) $\alpha_g {\cal A}(D) = {\cal A}(gD)$ (relativistic covariance: for any $g=(\Lambda, a) \in {\cal P}_+^\uparrow$, where
$gD = \Lambda D +a$).

Now, all one has to do, in order to apply the GNS construction and arrive at a quantum field theory, is to pick a state,
and usually this is chosen as the vacuum state, which, because of the cyclicity property, encodes a vast amount of information for the physical system.

\section{The Yang-Mills theory}

The self-interacting, spin-1, massless particle is described by the 
non-Abelian, Yang-Mills gauge theory, with coupling $g_c$, and  gauge group $G_c$, with generators $T^a$, structure constants $f^{abc}$,
gauge field $A_{\mu}= T^a A^a_{\mu}$ and
 $F^a_{\mu\nu}=\partial_{\mu}A^a_{\nu}-\partial_{\nu}A^a_{\mu} - g_c f^{abc}A^b_{\mu}A^c_{\nu}$.
The theory is gauge invariant, with
$A_{\mu}\rightarrow\omega A_{\mu} \omega^{-1} + \frac{i}{g_c}\omega\partial_{\mu} \omega^{-1}$,
 under the local gauge transformation
$\omega(\alpha) = e^{iT^a\alpha^a (x)} \in G_c$.

After imposing Gauss's  law with an auxiliary field (Lagrange multiplier)
$\lambda=\lambda^a T^a$, transforming covariantly as $\lambda\rightarrow\omega\lambda\omega^{-1}$,
and a set of appropriately modified Feynman rules, the gauge-invariant (although not manifestly Lorentz-invariant)
effective action
\begin{equation}
S_{\rm  eff}= \int \frac{1}{2}{E_i^a}E_i^a -\frac{1}{2}{B_i^a}B_i^a +
                        \lambda^a\, {D_i} {E_i^a} + U(\lambda),
\label{eff}
\end{equation}
was derived
(written in terms of the chromo-electric and -magnetic fields
$E^a_i = F^a_{0i}, B^a_i=-\frac{1}{2}\epsilon^{ijk}F^a_{jk}$)
with
\begin{equation}
U(\lambda)= \frac{(\alpha_s C_2)^2}{4}\lambda^4 \left(\ln\frac{\lambda^2}{\mu^2}-\frac{1}{2}\right),
\label{cw}
\end{equation}
($\alpha_s = g_c^2/4\pi$, $\lambda^2=\lambda^a \lambda^a$, $f^{acd}f^{bcd}=C_2\, \delta^{ab}$) a generated Coleman-Weinberg effective potential term, that appears inverted in the effective action. 
Essentially, what happens is that there are some missing terms involving the Coulomb interaction, that can be resummed to this term.
The equations of motion are 

\begin{equation}
D_i E_i^a = -\frac{\partial U}{\partial \lambda^a},
\label{eq1}
\end{equation}
\begin{equation}
D_i^2 \lambda^a = D_i E_i^a,
\label{eq2}
\end{equation}
\begin{equation}
D_0 \, E_i^a =  (D\times B)_i^a + D_0 D_i \lambda^a.
\label{eq3}
\end{equation}
The inverted effective potential,  $-U$, has a local minimum at $\lambda=0$, and a global maximum
at $\lambda^2=\mu^2$. However, because of the presence of gauge and kinetic terms, the analysis of \cite{dkm2}
showed that they are both stable, classically and quantum mechanically, vacua.
$\Omega_0$, with $\lambda = 0$, is the perturbative vacuum, with the usual Coulomb interaction,
and $\Omega_\mu$, with $\lambda^2 = \mu^2$, is the confining vacuum
(in the $A_0=0$ gauge, the vacuum $\Omega_\mu$ consists of time-independent, covariantly constant
configurations, $\lambda(\vec{x})= \omega(\vec{x}) \, \bar{\lambda}\, \omega(\vec{x})^{-1}$,
with $\bar{\lambda}$ a fixed adjoint vector $\bar{\lambda}^2=\mu^2$,  $A_i=\frac{i}{g_c}\omega(\vec{x}) \, \partial_i\, \omega(\vec{x})^{-1}$, $F_{\mu\nu}=0$).

Both vacua admit instanton solutions ($\vec{E}= \pm \vec{B}$), but since at most physical situations there are transitions between the two vacua, it was argued that the strong-CP violating, $\theta$ parameter, of Yang-Mills is zero.

The equations of motion also admit a soliton solution of the equation
\begin{equation}
\nabla^2 \lambda^a =-\frac{\partial U}{\partial \lambda^a}.
\label{sol}
\end{equation}
with $A_i=0, B_i=0, E_i =-\partial_i A_0 = \partial_i \lambda$.
Classical stability was demonstrated for all solutions of the equations of motion, including the soliton. It has a finite total energy (mass) and it connects the two vacua; in physical backgrounds with finite temperature and/or fermion density, the solitons mediate the transitions between the vacua.

The vacuum energy density of $\Omega_0$ is zero, and the vacuum energy density of $\Omega_\mu$ is positive (equal to $-U(\mu^2)$). 
The canonical formalism described in \cite{dkm2} shows that the energy (Hamiltonian) is given by
\begin{equation}
H =\int d^3x \left( \frac{1}{2} E_i^a E_i^a + \frac{1}{2} B_i^a B_i^a - U\right).
\label{h}
\end{equation}
In general, the effective action is not Lorentz-invariant, and there is no well-defined energy-momentum tensor. However, the two vacua do not decay, and at the vacua there is a well-defined energy-momentum tensor, 
\begin{equation}
\Theta^{\mu\nu}=\Theta^{\mu\nu}_{\rm YM} -  \eta^{\mu\nu} U
\label{emt}
\end{equation}
where
$\Theta^{\mu\nu}_{\rm YM}$ is the usual energy-momentum tensor
for perturbative Yang-Mills ($\Theta^{\mu\nu}_{\rm YM}=F^{a \mu \rho}{F^{a \nu}}_\rho -\frac{1}{4} \eta^{\mu\nu}F^{a\lambda\rho}F^a_{\lambda\rho})$.

Finally, because of the presence of a non-zero, $\lambda^2 = \mu^2$,  condensate in the confining vacuum, the confining interaction was demonstrated, with a linearly rising interaction energy between two color sources at a distance $r$, 
proportional to $C_2 \, \alpha_s \,\mu^2 r$.

In order to apply the axiomatic formalism of the previous Section in the Yang-Mills case, one first notices that the auxiliary field $\lambda$
commutes with all other fields of the theory (the canonical formalism, in the $A_0=0$ gauge,
implies the usual commutation relations between the canonical variables, $A_i$ and $E_i$).

The center of the algebra of the observables, therefore, is not trivial; 
it contains, besides unity, the gauge-invariant $\lambda^2$.
According to the discussion of the previous Section,  the GNS vacuum representation
$\pi_\varphi$, where $\varphi$ is the vacuum state,
splits into eigenspaces with $\lambda=0$ and $\lambda^2=\mu^2$, that is,
$\Omega_0$ and $\Omega_\mu$.

The Hilbert space of the theory is $H_0\oplus H_\mu$, and the vacuum state is mixed
\begin{equation}
\varphi = \rho_0 |\Omega_0><\Omega_0| + \rho_\mu |\Omega_\mu><\Omega_\mu|,
\label{v1}
\end{equation}
with positive constant coefficients, $\rho_0 +\rho_\mu=1$. Accordingly, vacuum expectation values
of the observables, $Q$, are given by
\begin{equation}
\varphi(Q) = \rho_0 <\Omega_0, Q \,\Omega_0> + \rho_\mu <\Omega_\mu, Q\, \Omega_\mu>.
\label{v2}
\end{equation}
Observables are the gauge- and Lorentz- invariant operators of the theory, and they obviously have completely different effects on the two vacua. The notation was simplified in (\ref{v2}); formally, one has $\pi_\varphi (Q) = (\pi_\varphi (Q)_0, \pi_\varphi (Q)_\mu)$, where the parentheses denote the components of the
direct sum. Cyclicity of the vacuum implies that
$(\pi_\varphi (Q)_0 \,\Omega_0, \pi_\varphi (Q)_\mu \, \Omega_\mu)$,
or $(Q \,\Omega_0,  Q \, \Omega_\mu)$ with less notation, is dense in $H_0\oplus H_\mu$.

The symmetries of the theory are also split in the two Hilbert spaces. Here, I will only consider
translation symmetry,  that is the subgroup of the Poincare group with $(\Lambda, a) = ({\rm\bf I}, a)$,
and denote the corresponding unitary operator on the GNS Hilbert space as $T_\varphi (a)$. This is also split
in the two Hilbert spaces, and there is an additional energy ``gap'', but for all states in the confining vacuum.
This is similar to old phenomenological, ``bag model", pictures of confinement.

In the next Section, I will elaborate more on the translation symmetry, and the cluster  decomposition, but before proceeding there, I would like to clarify some possible confusion regarding the superselection sectors that may
also appear in reducible GNS constructions. These are not decompositions of the vacuum representation, they 
appear whenever there is a conserved quantity (charge) of a continous of discrete symmetry, they also appear
in QED, in QCD, wherever one can properly define a conserved, gauge-invariant charge. They are decompositions of a general reducible GNS 
construction into sectors with different charge (there are some lower-dimensional cases with different vacua, with the same energy density, connected with a discrete symmetry, but these are also irrelevant to the present work).

The derivation of confinement, and the associated splitting of the vacuum state shown here, are purely due to the dynamics, the self-interaction of the spin-1 field. Other properties of the strong interaction, asymptotic freedom, symmetries, anomalies, etc., have no relation to confinement and the vacuum state (no other than being derived from the same theory).

\section{Vacuum state and clustering property}

Here, I will describe some more properties of a general vacuum state, following \cite{qft3}, which should be consulted for the proofs that are not given here, as well as the issues of convergence. Then, I will note the 
differences between a pure and mixed vacuum state with respect to clustering properties, and make contact with 
the solution of Yang-Mills.

The expression of the symmetry of the proper, orthochronous, Poincare group, ${\cal P}_+^\uparrow$, on 
a representation, $\pi$, of an algebra, ${\cal A}$, is generally given by
\begin{equation}
\pi(\alpha_{(a,\Lambda)} (Q))= U(a, \Lambda) \pi (Q) U(a, \Lambda )^*,
\end{equation}
for any $Q\in {\cal A}$, where $(a, \Lambda) \in {\cal P}_+^\uparrow $ is a general Poincare transformation (consisting of a translation $a$, and a Lorentz transformation $\Lambda$), $\alpha$ is the action of the symmetry on the algebra, ${\cal A}$,
and the unitary $U$ is the symmetry operator on the Hilbert space of the representation.
The action of translations, where $\Lambda = {\rm \bf I}$, will be denoted by $\alpha_a$ and the unitary $T(a)$.
The spectral decomposition of the unitary operator of a translation is
\begin{equation}
T(a)=\int e^{i a\cdot p} \, E(d^4 p)
\end{equation}
for four-momenta $p=(p^0, \vec{p})$, where $E$ is a projection-valued measure on momentum space.
Then, for any vector $\xi$ of the Hilbert space, the support of the measure $||E(d^4 p)\xi||^2$ is the energy-momentum spectrum of $\xi$. The following definitions and theorems can be found in \cite{qft3}.

(i) If $\tilde{M}$ is the four-dimensional momentum space, $\Delta$ a compact subset of $\tilde{M}$,
and $Q$ an element of the algebra ${\cal A}$, then
\begin{equation}
Q(g)=\int \alpha_x(Q) g(x) d^4 x,
\end{equation}
is an operator that shifts the momentum by $\Delta$
if 
\begin{equation}
\tilde{g}(p)=\int g(x) e^{i p \cdot x} d^4 x
\end{equation}
satisfies
${\rm supp} \,\tilde{g} \subset \Delta$ (the support of a function being the set of its non-zero points).

(ii) If the energy-momentum spectrum of a vector $\xi$ of the Hilbert space is in $F\subset \tilde{M}$,
the energy momentum spectrum of $Q(g) \xi$ is in $F+\Delta$.

(iii) If $\hat{e}$ is any positive timelike unit vector of $\tilde{M}$, and 
$\tilde{M}(\hat{e})_- =\{p\in\tilde{M}, p\cdot \hat{e} < 0\}$,
an operator $Q(g)$ will be called an energy decreasing operator
if ${\rm supp}\,\tilde{g}\subset \tilde{M}(\hat{e})_-$.

(iv) The vacuum is defined as a state $\varphi$, such that $Q\in {\rm ker} \,\varphi$ for any 
energy decreasing operator.
Here, as in (\ref{ker}),
${\rm ker}\,\varphi =\{Q\in {\cal A}: \varphi(Q^* Q) =0\}$, and the vacuum is defined as a state that is
stable under any local perturbation.

(v) The vacuum state is translationally invariant, that is $\varphi(\alpha_x Q)=\varphi(Q)$ for any $Q\in {\cal A}$,
and any $x$ in $M$.

(vi) On the GNS representation space of the vacuum state
\begin{equation}
\pi_\varphi (\alpha_x Q) = T_\varphi (x) \pi_\varphi (Q) T_\varphi^*, \,\,\,\,\,\,
T_\varphi(x) \, \Omega_\varphi = \Omega_\varphi,
\end{equation}
and 
\begin{equation}
T_\varphi (a) = \int e^{i p\cdot a} E_\varphi (d^4p ),
\end{equation}
where the spectral measure, $E_\varphi$, has support on the positive light-cone of $\tilde{M}$
($p^2 \ge 0, p^0  \ge 0$).

From now on, we focus on the GNS representation space of the vacuum state, for which the same observation holds for irreducibility that was described in Sec.~2. Namely, irreducibility is equivalent to triviality of the center, 
$\pi_\varphi({\cal A})' \cap \pi_\varphi({\cal A}) $.
As was described in Sec.~3, in the case of Yang-Mills theory, the center is non-trivial, since it contains the
gauge- and Lorentz- invariant operator $\lambda^2$.

Irreducibility of the vacuum state is also equivalent to the condition that there is a unique translation invariant vector, that is, any translation invariant vector
is proportional to $\Omega_\varphi$ \cite{qft3}. This was also shown not to hold in the Yang-Mills case, since there are two
distinct translationally invariant vectors, $\Omega_0$ and $\Omega_\mu$.

The GNS vacuum representation space splits in the direct sum of two components,
the eigenspaces of $\lambda^2=0$ or $\mu^2$,
 the perturbative and the confining vacuum, as described before, and so do the various operators and symmetries, for example,
\begin{equation}
T_\varphi (a) = \left(\int e^{i p_{(0)}\cdot a} E^{(0)}_\varphi (d^4p ), 
                              \int e^{i p_{(\mu)}\cdot a}  E^{(\mu)}_\varphi (d^4p ) \right),
\end{equation}
where, the $(0)$ and the $(\mu)$ quantities that correspond to the $\Omega_0$ and $\Omega_\mu$ vectors, are calculated with (\ref{h}) or (\ref{emt}), and are different in each vauum; in the confining vacuum, there is a ``mass gap'', but for all states at the corresponding Hilbert space, $H_\mu$.

Finally, irreducibility is also equivalent to a general clustering property:
\begin{equation}
\lim_{s\to\infty} \varphi (Q_1 \alpha_{sx} Q_2) = \varphi(Q_1) \varphi(Q_2),
\label{cl}
\end{equation}
where $x$ is any spacelike vector in $M$, and $Q_1$, $Q_2$ any elements of ${\cal A}$.

Indeed,  first assume that (\ref{cl}) holds: then, if we denote the projection operator to
$\Omega_\varphi$ by $E_\Omega$, and the projection operator to all translationally invariant vectors
by $E_0$,
(\ref{cl}) can be written as
\begin{equation}
\lim_{s\to\infty} < \Omega_\varphi, \pi_\varphi (Q_1) T_\varphi (sx) \pi_\varphi (Q_2) \Omega_\varphi>
= <\Omega_\varphi, \pi_\varphi (Q_1) E_\Omega \pi_\varphi (Q_2) \Omega_\varphi >.
\end{equation}
Since the $Q_{1,2}$ are arbitrary, and $\pi_\varphi (Q) \Omega_\varphi$ are dense in the Hilbert space,
it follows that $\lim_{s\to\infty} T_\varphi (sx) = E_\Omega$, and if we multiply this relation by $E_0$
we get $E_0 = E_\Omega$, thus, (\ref{cl}), clustering, implies uniqueness of a translationally invariant vacuum.

Conversely, uniqueness of a translation invariant vacuum (irreducibility of the vacuum state) implies
clustering. Indeed, first notice that, as $s\rightarrow\infty$, $Q_1$ and $\alpha_{sx} Q_2$, with spacelike $x$,
eventually are in spacelike separated regions, and commute, by the axiom of causality. Since the center is trivial,
the limit of $\alpha_{sx} Q_2$ exists and is also trivial, say $c {\rm \bf I}$, with a complex constant $c$. By translational invariance
$\varphi(\alpha_{sx} Q_2)=\varphi(Q_2)=\varphi(c {\rm \bf I}) = c$, hence we get the clustering property (\ref{cl}).

On the other hand, for a reducible vacuum state, clustering does not necessarily hold:
let us consider the case of Yang-Mills, described before, where one has the simplest, perhaps, case of reducibility,
with a mixed vacuum state consisting of two distinct translationally invariant vectors, $\Omega_0$ and $\Omega_\mu$.

As was explained, the Hilbert space of the theory is $H_0\oplus H_\mu$, and the vacuum state is mixed
\begin{equation}
\varphi = \rho_0 |\Omega_0><\Omega_0| + \rho_\mu |\Omega_\mu><\Omega_\mu|,
\end{equation}
$\rho_0, \rho_\mu \in [0, 1]$, $\rho_0 +\rho_\mu=1$. Vacuum expectation values
of the observables, $Q$, are given by
\begin{equation}
\varphi(Q) = \rho_0 <\Omega_0, Q \,\Omega_0> + \rho_\mu <\Omega_\mu, Q\, \Omega_\mu>
\equiv \rho_0 <Q>_0 +\rho_\mu <Q>_\mu.
\end{equation}
Since the center now is not trivial, but spanned by ${\rm \bf I}$ and $\lambda^2$,
the previous limit of $\alpha_{sx} Q$ is not well-defined. In fact, different sequences $\alpha_{s_n x} Q$,
with $s_n \rightarrow\infty$, may have different accumulation points,
of the form $c_0 {\rm \bf I} + c_{\mu} \lambda^2$, with complex constants $c_0, c_\mu$.
If we pick one such point, then the limit
of $\varphi(Q_1 \alpha_{sx} Q_2)$ is $\varphi(c_0 Q_1 + c_\mu \lambda^2 Q_1)=
c_0(\rho_0  <Q_1>_0 +\rho_\mu <Q_1>_\mu) +\rho_\mu c_\mu \mu^2 <Q_1>_\mu$.
On the other hand, $\varphi(\alpha_{sx} Q_2)=\varphi(Q_2)=\varphi(c_0{\rm\bf I}+c_\mu \lambda^2)=
c_0 + \rho_\mu c_\mu \mu^2$, and $\varphi(Q_1) =\rho_0 <Q_1>_0 +\rho_\mu <Q_1>_\mu$,
and it is easy to see that the clustering property, in general, does not hold.

Obviously, what happens physically, is that, when two sources that are initially in one vacuum vector are pulled apart, they 
may remain in the same vacuum, or jump in the other vacuum vector, or in a mixed vacuum state.
Hence, the clustering property does not hold, and the restriction of the attenuation of the interaction energy is also lifted.

Generally, in axiomatic field theory, under the assumption of a unique, pure vacuum state, one goes on to prove stronger
clustering properties than (\ref{cl}), namely, that the limit to a product state is reached at large spacelike separations
bounded by a Coulomb or Yukawa interaction.
It should be stressed that the results of axiomatic field theory are non-perturbative; if the clustering property holds, it is not possible to obtain a linearly rising, confining, potential  energy.

So, the existence of a mixed vacuum state is necessary for the theory of the strong interaction.
On the other hand, between the two vacua described here, there is also an energy density difference. This implies that 
a Lagrangian description of the system is not possible. Indeed, if there was such a Lorentz-invariant Lagrangian (in place of (\ref{eff})), 
it would be possible to build a local configuration, using the gauge- and Lorentz- invariant operators of the Lagrangian (not necessarily satisfying the equations of motion) that perturbs one vacuum, creates a small bubble of the other vacuum, and thus lowers the energy, which is not possible by the discussion and definitions in the beginning of this Section.
In our case, (\ref{sol}) creates a small bubble of the confining vacuum in the perturbative vacuum and increases the energy, but the operators involved are not Lorentz-invariant. So, the solitons described, and their combinations, do not belong to any of the Hilbert spaces, $H_0$ or $H_\mu$. They are mediators
of the phase transitions between the two spaces (phases) in physical situations, with a background temperature and matter density.

\section{Discussion}

The absence of a Lagrangian description explains, in a loose sense, the success of many older phenomenological models for confinement, and various other approaches, with physical insight but lack of some conventional properties of quantum field theory, such as renormalizability, for example. Besides the relation of the present work with the bag model \cite{bag}, a connection with the Nambu-Jona-Lasinio model \cite{njl} was also established in \cite{dkm2}. 
Furthermore, other works with interesting proposals of effective actions can be mentioned, such as \cite{kogut}, that explore the dielectric properties of the vacuum, and
\cite{pagels}, that investigate possible gauge field condensates.

More connections of the present work with these and other studies can be investigated further by considering additional terms in the effective action,
such as, for example, a wavefunction renormalization factor, $Z(\lambda)$, that multiplies the gauge and Gauss law terms in (\ref{eff}), and can be related to the dielectric and diamagnetic properties of the vacuum. It should be noted, however, that, when the values of $Z$ are considerably different than unity, the perturbative calculations will inevitably break down.
At finite temperature and fermion density, corrections to the effective potential term, $U(\lambda)$, in (\ref{eff}) can also be calculated. Other, intermediate vacua, may consequently arise, and transitions between the two original vacua are then possible, as one can investigate via well-known methods; some more arguments to that effect were also made in \cite{dkm2}.
The fact that the finite-temperature phase transition in Coleman-Weinberg models is typically of the first order, already gives an encouraging prediction for the order of the deconfinement phase transition. It should be noted, however, that, at finite temperature and density backgrounds, there are more terms that may be important in the effective action. The present work, which concludes and axiomatically justifies the results of \cite{dkm1} and \cite{dkm2}, deals with pure Yang-Mills theory at zero temperature.
Here, the presence of the two stable vacua, and the resulting mixed vacuum state, is essential for the description of the quantum properties of the theory.

\section{Comments}

The postulate of a unique, translation invariant vacuum (a pure vacuum state) is unfitting a theory of strong interactions. It is equivalent to the cluster decomposition property, and was originally used in axiomatic field theory with a view to describe scattering and asymptotic states for accelerator setups. However, the strong interaction, in the confining phase, does not admit asymptotic states. Colored sources are dynamically confined by a linearly rising interaction at this scale, and the entire spectrum has a ``mass gap''.

As far as a related, mathematical problem is concerned \cite{clay}, it should be clear from these results, that the statement of the problem itself is wrong, or at least incomplete, in that it assumes a pure, unique vector, vacuum state. Thus stated, the problem is a red herring; it cannot have a solution with a physical theory that corresponds to the strong interaction.

Once the picture of the spin-1 interaction is better understood, one would like to apply a similar approach to the spin-2 case, which was, in fact, the initial motivation of \cite{dkm1}. One can immediately notice that the possibility of another vacuum with positive energy density can be related to an inflationary, cosmological, de~Sitter phase. Thus, an inflationary period may arise, purely based on the dynamics of the spin-2 interaction, without the need of an inflaton field.
Some more, also naive observations, regarding the existence of a mixed vacuum state and the breakdown of the clustering property, are the presence of a de~Sitter horizon, and the presence of a polynomial interaction, besides the Coulomb term, for a static observer in a vacuum with a cosmological constant (${\sim H^2 r^2}$, for a distance $r$, where $H$ is the Hubble constant).

Provided that a renormalisable interaction (``vertex'') can be identified, besides the quadratic Fierz-Pauli term, and the constraints are treated, one would like to arrive at an expression like (\ref{eff}), with different, stable vacua. Then, resummations of the effective equations of motion in each vacuum \cite{deser}, would lead to different phases: some would correspond to ordinary gravity (general relativity with a zero cosmological constant), others to an inflationary phase, others may even be non-interacting, or ``modified'' gravity vacua.
It should be clear  that the picture that emerges is fundamentally, conceptually different than any inflationary, eternal inflation, or multiverse scenario. Apart from the absence of additional inflaton fields, what is described here are different
phases, different vacua of the same spin-2 field, that correspond to different effective theories, and may exist at the same cosmic fluid, in other cosmological regions or scales.

However, the problem of finding a renormalizable ``vertex'', a dimensionless coupling constant, in order to apply the techniques of ordinary quantum field theory, is still not solved.
It is possible that, much like the full Yang-Mills case does not admit a Lagrangian description, for the spin-2 interaction, even the ``vertex'' term may not admit a Lagrangian description. Hence, the formalism of algebraic quantum field theory seems necessary.

%\vspace{0.5in}

\end{document}